\documentclass[paper=letter, american, fontsize=11, DIV=10]{scrartcl}
\usepackage{times}
\usepackage[T1]{fontenc}
\usepackage[utf8]{inputenc}
\usepackage{babel}
\usepackage{amsmath}
\usepackage{amssymb}
\usepackage{amsthm}
\usepackage{xfrac} 
\usepackage{dsfont}
\usepackage{verbatim}
\usepackage{epigraph} 
\usepackage{graphicx}	

\usepackage[authoryear,round]{natbib}
\usepackage[indentfirst=false,font=normal]{quoting}
\usepackage{latexsym}
\usepackage{authblk} 
\usepackage[unicode=true,bookmarks=true,bookmarksnumbered=false,bookmarksopen=false, breaklinks=false,pdfborder={0 0 0}, backref=false,colorlinks=false]{hyperref}   

\renewcommand{\vec}{\boldsymbol}

\newtheorem*{theorem*}{Theorem}

\begin{document}
\pagenumbering{roman}

\title{Reviving Frequentism\thanks{Accepted for publication in \emph{Synthese}.}}

\author{Mario Hubert}
\affil{California Institute of Technology\\
Division of the Humanities and Social Sciences}
\date{December 31, 2020}

\maketitle
\begin{abstract}
Philosophers now seem to agree that frequentism is an untenable strategy to explain the meaning of probabilities. Nevertheless, I want to revive frequentism, and I will do so by grounding probabilities on typicality in the same way as the thermodynamic arrow of time can be grounded on typicality within statistical mechanics. This account, which I will call \emph{typicality frequentism}, will evade the major criticisms raised against previous forms of frequentism. In this theory, probabilities arise within a physical theory from statistical behavior of almost all initial conditions. The main advantage of typicality frequentism is that it shows which kinds of probabilities (that also have empirical relevance) can be derived from physics. Although  one cannot recover all probability talk in this account, this is rather a virtue than a vice, because it shows which types of probabilities can in fact arise from physics and which types need to be explained in different ways, thereby opening the path for a pluralistic account of probabilities. 
\end{abstract}

\newpage

\tableofcontents

\newpage

\section{Introduction}
\pagenumbering{arabic}

Frequentism is dead. This seems to be the consensus among contemporary philosophers. A recent textbook on the philosophy of probabilities phrases it this way:

\begin{quotation}
Although the frequency view remains popular outside philosophy -- e.g.\ among statisticians -- it is not the subject of much, if any, active research. \citep[][p.\ 112]{Rowbottom:2015aa}
\end{quotation}
Frequentism may be useful for all practical purposes for statisticians, although it does not convey the true meaning of probabilities, since philosophers have successfully exposed the underlying unsurmountable problems. Therefore, active research for developing frequentism has been discontinued. \citet{Hajek:1996aa} prominently debunked finite frequentism; a decade later followed his criticism of hypothetical frequentism \citep{Hajek:2009aa}. Recently, \citet{Caze:2016aa} agreed that any version of frequentism is doomed to fail, at least in providing a comprehensive understanding of probabilities. 

I think we can breathe life back into frequentism and develop it into a serious account of probabilities. I intend to defend frequentism against these criticisms and modify it in such a way that it incorporates elements of finite frequentism,  hypothetical frequentism, and the classical interpretation of probabilities. I will call this account \emph{typicality frequentism}, which defines, in brief, probabilities as typical long-term frequencies based on the law of large numbers.

Typicality has been developed within Boltzmann's reduction of thermodynamics to statistical mechanics, but the scope of this notion is not particularly tied to statistical mechanics \citep{Wagner:2020aa}. \citet{Wilhelm:2019aa} recently showed how typicality explanations work in general by connecting them with Hempel's deductive–nomological model. Ideas along these lines to derive probabilities from typicality as special kinds of frequencies have been presented by \citet{Maudlin:2007ac,Maudlin:2019ab} and sketched in the literature on Boltzmann's statistical mechanics and the de Broglie--Bohm quantum theory \citep[for a brief overview, see][]{Goldstein:2012aa}, but there has been no work contrasting this kind of frequentism with the traditional theories of frequentism in order to establish typicality frequentism as a serious alternative theory in its own right. 

In my opinion, the biggest methodological error made by the forefathers of frequentism, like \citet{Reichenbach:1949aa}, \citet{Venn:1888aa}, and \citet{Mises:1957aa}, was to interpret probabilities as frequencies from empirical behavior: they started with how we talk about probabilities and tried to underly an interpretation in terms of frequencies that supports their empiricism \citep[][Ch.\ 5]{Gillies:2000aa}. Instead, I propose a strategy from a physical theory to probabilities: starting with a deterministic fundamental physical \emph{theory} and analyze how this theory introduces probabilities from statistical behavior. Then we may recover how our general use of probabilities is backed up by physics. But, as it turns out, some ordinary ways of talking about probability will not be recovered within this approach. To account for these, we are free to introduce another, complementary interpretation of probability---becoming pluralists about probability. The method I will be using to define probabilities is the same statistical method that has been used to justify the thermodynamic arrow of time in statistical mechanics or the arrow of time in electrodynamics \citep{North:2003aa}.

\section{Typicality Frequentism}

The idea behind typicality frequentism is to apply the tools from statistical mechanics to explain how probabilities arise from deterministic physical dynamics. \citet{Maudlin:2019ab} is confident about this strategy, 
``The challenge of deriving probabilities---or quasi-probabilities, probabilities with tolerances---from an underlying deterministic dynamics can be met. Typicality is the conceptual tool by which the trick is done.'' An important predecessor of typicality frequentism, apart from the different versions of frequentism, is the theory of probability by Johannes von Kries, laid out in his \emph{Principien der Wahrscheinlichkeitsrechnung} (\citeyear{Kries:1886aa}, \emph{engl.} Principles of Probability Theory). As von Kries's view seems to be best characterized as objective Bayesianism \citep[see p.\ 16 of the introduction by Eberhardt and Glymour in][]{Reichenbach:2008aa} or as a predecessor of the logical interpretation \citep{Fioretti:2001aa}, subjective and objective aspects are intertwined. For my purpose, I want to lay out in more detail the objective parts of von Kries's account, because they contain some essential features of typicality frequentism, although von Kries criticized the frequentist theories at his time \citep[][section 3]{Zabell:2016ab}.  

Influenced by the physics of the 19\textsuperscript{th} century, it was important to von Kries to distinguish between laws of nature and initial conditions \citep[][]{Pulte:2016aa}. Unlike Laplace, who reduced probability to incomplete knowledge of the initial conditions, von Kries built up probabilities from objective \emph{variations} of initial conditions, and he called the sets of admissible initial conditions ``Spielräume,'' which are best translated as ``sets of possibilities''.\footnote{Eberhard and Glymour call them ``sets of ur-events'' because they are the irreducible basis for von Kries's probabilities \citep[see][Introduction, section 4.2]{Reichenbach:2008aa}.} More quantitatively, von Kries defined the probability for an event $E$ in the following way \citep[see][section 5, for this reconstruction]{Pulte:2016aa}. Let us say that the event $E$ is brought about by the set of initial conditions $C$ (given certain laws of nature) and the set of initial conditions that do \emph{not} bring about $E$ is $C^*$ ($C$ would be then the ``Spielraum''  or set of possibilities for $E$). The probability $p$ for $E$ is then defined, in the Laplacian sense, as the quotient of the favorable initial conditions leading to $E$ over all possible initial conditions by measuring the Spielraum and its complement by an appropriate measure $m$:
$$
p:=\frac{m(C)}{m(C)+m(C^*)}.
$$
Although the objective aspect of probabilities mentioned here comes from the initial conditions of the physical process, it is not a frequentist account. Moreover, this reconstruction of von Kries's theory may incline us to think that the Spielräume are unique and always tied to a physical theory, but von Kries was in this respect more a pragmatist and sometimes even  a skeptic \citep[see the discussion in][section 5]{Pulte:2016aa}. Depending on the knowledge of the agent building a probabilistic model, the space of possibilities may change and may not be a space of initial conditions of a physical theory but rather a more generalized sample space; thus, the measure $m$ may not be unique either. 

 I share von Kries's intuition to reduce probabilities to certain basic events, but I endeavor a more objective account of probabilities always embedded into physics and using only the tools of physics in defining probabilities. I, therefore, propose that physics offers a unique space from which to derive probabilities as typical long-term frequencies. This is the fundamental space of physics, like \emph{phase space} or \emph{configuration space} (depending on the physical theory). Here, I agree with the method of arbitrary functions  \citep[or more adequately named the \emph{range account of probabilities} by][]{Rosenthal:2016aa}, which can be regarded as a modern elaboration of von Kries's theory. But I deviate from the range account by incorporating typicality as a central notion to define probabilities; in a similar fashion, it is possible to explain the thermodynamic arrow of time as arising from generic initial conditions of the universe. 

\subsection{Typicality and the Arrow of Time}

Scrambled eggs never unscramble, a shattered vase never reassembles itself, and ice cubes never un-melt. Although our basic physical laws are time-reversal invariant, that is, the time-reversed process of a physically possible process is also physically possible, such time-reversed processes are not observed. Boltzmann proposed a solution to this problem by distinguishing microscopic from macroscopic behavior and systems with few degrees of freedom from systems with many degrees of freedom. It is the microscopic behavior that is time-reversal invariant, and one only finds directed processes on the macroscopic level, when systems have many degrees of freedom. If we have a sequence of photos showing a system of few degrees of freedom, like two rotating molecules, we would not be able to distinguish forward from backward behavior, but if we have a sequence of photos of a glass bottle thrown on the ground, we would distinguish one direction as the true one.

Boltzmann gave an explanation in terms of statistics why such behavior is not observed. The short answer is due to the many degrees of freedom of a macroscopic process: one had to finely orchestrate all the many microscopic states of the particles constituting a macroscopic object in order to yield a time-reversed process. If we don't interfere this meticulously (and in most cases we cannot do so), then a familiar directed process comes about.  In other words, almost all initial conditions of a macroscopic system yield the familiar directed processes; only very special initial conditions yield the reversed process. So given broken glass on the floor, there are many more states of particles constituting the pieces of glass such that these pieces remain on the floor than those states that would converge the pieces into a brand-new bottle.  

This behavior can be phrased by means of typicality: a physical behavior is called \emph{typical}, if almost all initial conditions yield this behavior \citep[see, for instance,][]{Goldstein:2001aa,Lebowitz:2008aa,Myrvold:2016aa,Volchan:2007aa}. And a physical behavior is called \emph{atypical}, if almost none of the initial conditions yield this behavior. So, it is typical that broken glass remains broken, and it is atypical that scrambled eggs unscramble. 

\begin{figure}[ht]
\centering
\includegraphics[width=8cm]{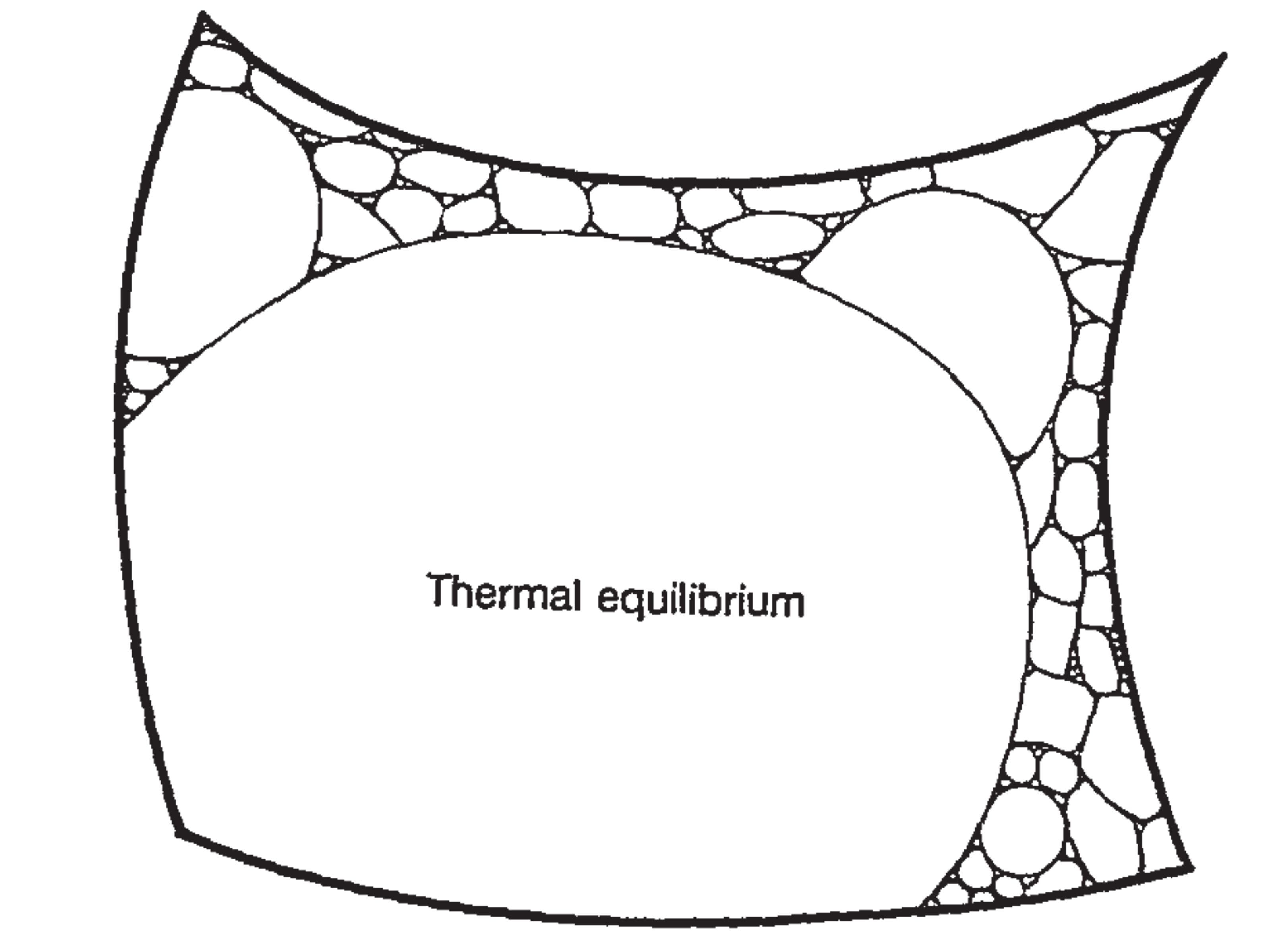}
\caption{Clusters in phase space according to thermodynamic macrostates and the measure of typicality, as depicted by Roger \citet[][p.\ 402]{Penrose:1989aa}. Thermal equilibrium is by far the biggest macrostate in phase space.}
\label{fig:phase-space-cluster}
\end{figure}

Boltzmann's ideas on the irreversibility of physical processes have recently experienced a renaissance among philosophers in which the notion of typicality has become central \citep[see, for instance,][]{Barrett:2017aa,Frigg:2009aa,Lazarovici:2015aa,Lazarovici:2019ab}. 
The notion of typicality, as we introduced it, is still too imprecise for quantitative use in physics. As \citet{Wilhelm:2019aa} rightly emphasizes, there are many ways to formalize ``almost all.'' The right way to do so in statistical mechanics is by means of a measure over phase space. Phase space is constructed from a set of particles in three-dimensional space.\footnote{For simplicity's sake and to be as close to Boltzmann's reasoning as possible, I assume Newtonian mechanics as the microscopic theory.} Consider $N$ particles in three-dimensional space (if we have a realistic macroscopic body, $N$ is of the order of Avogadro's constant, that is,  approximately $10^{23}$). Since a particle is completely described by its position $\vec{x}$ and momentum $\vec{p}$, we can summarize the complete physical state of $N$ particles as a vector $\left( \vec{x}_1,\vec{p}_1,\vec{x}_2,\vec{p}_2,\dots,\vec{x}_N,\vec{p}_N\right)$, and this vector is one point in phase space, which has roughly $6\times10^{23}$ dimensions. So every point in phase space, each microstate, represents a set of $N$ particles with their precise positions and momenta. In order to get macrostates, one needs to divide phase space $\mathcal{P}$ into disjoint subsets, where each set represents a macrostate (see Figure \ref{fig:phase-space-cluster}). So a macrostate arises  from a map $M$ that assigns to every microstate $X$ a macrostate $M(X)$ corresponding to one of the subsets $\mathcal{P}_M\subseteq \mathcal{P}$ according to the partition---$M(X)$ is the macrostate of $X$ if $X\in\mathcal{P}_M$.

The tool that ultimately explains irreversible behavior is Boltzmann's definition of entropy assigned to every point in phase space:

\begin{equation}
\label{eq:entropy}
S_B(X):=k_B\ln \lvert \mathcal{P}_{M(X)}\rvert, 
\end{equation}
where $k_B$ is Boltzmann's constant and $\ln$ is the natural logarithm. The main part of Boltzmann's entropy is $\lvert \mathcal{P}_{M(X)}\rvert$, which deserves some elaboration. In order to measure the sizes of the subsets $\mathcal{P}_{M(X)}$, one needs to introduce a measure $\lambda$, which assigns a number to every such subset. Conventionally, if the system is finite, one normalizes the measure to $1$ such that the size of the entire phase space would be $1$. In the entropy formula,  $\lvert \mathcal{P}_{M(X)}\rvert$ denotes the size of $\mathcal{P}_{M(X)}$ according to the appropriate measure $\lambda$, that is, $\lvert \mathcal{P}_{M(X)}\rvert=\lambda \left( \mathcal{P}_{M(X)}\right)$. The only purpose of the measure $\lambda$ is to tell us which sets are big and which are small, in order to identify typical and atypical behavior; in this sense, it is \emph{a measure of typicality}. Since for real physical systems, like gases in a box or melting ice cubes, the phase space volume of thermal equilibrium has by far the largest volume according to the measure of typicality and so it has the highest entropy, we observe systems that are not in equilibrium (low entropy $S_B$) to reach equilibrium (high entropy $S_B$), whereas we do not see a system going from a high entropy state to a low entropy state, because the low entropy states are much smaller in phase space.

Moreover, if we zoom into the phase space region of a low entropy macrostate, a melting ice cube, for example, almost all microstates will move to a macrostate with higher entropy and ultimately to equilibrium. It is physically possible that a low entropy macrostate goes into another low entropy macrostate (by itself), but there are very few microstates within this macro region that do that. This is Boltzmann's explanation why we observe only one direction of a physical process and not the time-reversed process, although this behavior is physically possible according to the time-reversal fundamental laws. The symmetry is broken by a statistical argument, that is, by distinguishing those initial conditions that yield typical behavior from those that yield atypical behavior. 

In classical mechanics, one normally uses the \emph{Liouville measure}, a natural generalization of the standard \emph{Lebesgue measure} on three-dimensional space to phase space, as the measure of typicality. But in order to distinguish small sets from big sets, other measures would do the job as well. Indeed, every measure that is \emph{absolutely continuous} with respect to the Liouville measure will agree on the same physical behavior to be typical or atypical.\footnote{A measure $\mu$ is absolutely continuous with respect to a measure $\lambda$ (symbolically $\mu \ll \lambda$),  if all the null sets of $\lambda$ are null sets of $\mu$, that is $\forall X \left( \lambda(X)=0 \Rightarrow \mu(X)=0\right)$.} 
Moreover, there is a certain vagueness intended in the notion of typicality that is to be reflected in the mathematical formalization. The sets $A$ yielding typical behavior are those that have measure $1$ or close to one, that is, $\lambda(A)=1-\epsilon$, where $\epsilon$ is a very small number also depending on the application. Similarly, for atypical behavior where the relevant sets may have a measure $\lambda(B)=0+\delta$ for some small $\delta$, which depends on the specific application.\footnote{There has been a long debate to make Boltzmann's argument more mathematically and conceptually precise. For our purpose, we do not need to dive into these details \citep[see, e.g.,][]{Volchan:2007aa,Frigg:2009aa,Werndl:2013aa,Lazarovici:2015aa,Myrvold:2019ab}.} This will become important when we apply typicality and its mathematical formalizations to develop a new theory of frequentism.

\subsection{Probabilities as Typical Frequencies}

There are two steps to present the theory of typicality frequentism. First, I need to elucidate the role of random variables, in a way that differs from standard accounts of probability theory (section \ref{subsec:random-var}). In typicality frequentism, random variables are primarily used to bridge the gap between a \emph{physical} theory and the mathematics of probability theory. Second, this account of random variables is needed to interpret the law of large numbers in such a way to define probabilities as typical long-term frequencies (section \ref{subsec:prob-typ}).

\subsubsection{Random Variables and their Relation to Physics}
\label{subsec:random-var}

\begin{figure}[ht]
\centering
\includegraphics[width=8cm]{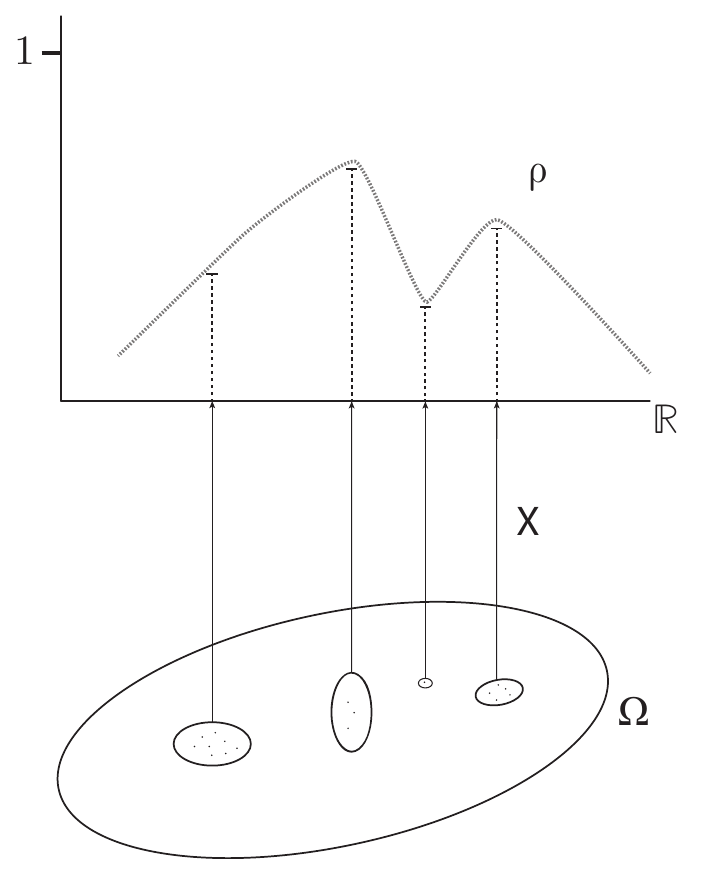}
\caption{The ingredients of a stochastic model and how they relate to each other. Random variables $X$ abstract from the sample space $\Omega$ by assigning every member of $\Omega$ a real number. Abstracting means that $X$ maps many elements in its domain to the same number. The image of $X$ gets assigned a number in the interval $[0,1]$, which measures the size of the sets that are mapped by $X$ to the same real number. It's important for typicality frequentism that all random variables are ultimately defined on  phase space, which is the fundamental sample space.}
\label{fig:random-var}
\end{figure}

Consider a box with $1000$ balls; the balls are either blue, green, or red. Let's say 500 balls are blue, 300 are green, and 200 are red; in other words, $50\%$ are blue, $30\%$ are green, and $20\%$ are red.  With this information we can build a simple stochastic model. The set of balls forms the \emph{sample space} $\Omega:=\{1,\dots,1000\}$. From this sample space, we can define a coarse-graining function $X:\Omega\rightarrow\{B,G,R\}$, which assigns to every ball a color B=blue, G=green, or R=red. Functions of this kind are usually (and unfortunately misleadingly) called \emph{random variables}. There is indeed nothing random about them; their only use is to abstract from the sample space, when one is interested in specific features of the members of the sample space. Next, one determines the \emph{distribution} of the random variable. This is a function $\rho_X:F(X)=\{B,G,R\} \rightarrow [0,1]$, such that $\rho_X(B)=0.5$, $\rho_X(G)=0.3$, and $\rho_X(R)=0.2$. This illustrates the standard way of building a \emph{probability space} (see Fig.\ \ref{fig:random-var}).\footnote{There are some subtleties when one generalizes this scheme to infinite sample spaces, like, forming a $\sigma$-algebra. These are treated in standard textbooks on probability theory and are not the focus of this paper.}

 The distribution $\rho$ is normally called a \emph{probability distribution}, for it assigns ``probabilities'' to certain sets of the sample space. But this would be putting the cart before the horse; at this stage, we do not have a theory of probabilities, just a certain recipe for building a mathematical model. This particular model of colored balls is conceptually very simple, because the numbers $50\%$, $30\%$, and $20\%$ are mere proportions of balls having the same color. Nevertheless, some work is to be done to interpret these numbers correctly as probabilities, as we will be doing in the next subsection, when I fully lay out typicality frequentism. 

The sample space can be in principle any kind of (mathematical) space, and in general no particular attention is paid to the sample space in textbooks, because in order to make correct predictions the images of the random variables and the probability distribution are sufficient. I want to go beyond a pragmatic attitude toward probability theory, although it is justified by its success in application, and derive probabilities instead from physical behavior. This is where von Kries's idea of sets of possibilities or ur-events comes in. He wanted to prove that probabilities can be derived from certain compositions of ur-events---random variables, in his account, are defined on these spaces. Although he intended an objective theory of probability, von Kries had to rely on a subjective element in order to justify that ur-events are equiprobable. This element is \emph{the Principle of Indifference}, which he advocated in the form of a principle of \emph{in}sufficient reason: Two events are equipossible if at the current state of knowledge there is no reason to consider one of the events more likely than the other \citep[][p.\ 15]{Reichenbach:2008aa}.

It is, however, possible to retain ur-events without this subjective ingredient. There is a distinguished sample space among all possible sample spaces, namely, phase space, on which a typicality measure can be defined, thereby erasing the principle of insufficient reason.\footnote{If one were to embed this discussion in quantum theory, one would need to replace phase space with configuration space.} I now make the following postulate: \emph{all random variables are ultimately defined on phase space, because all statistical patterns are determined by what happens in the fundamental physical space, which are governed by the laws of physics}. Hence, probability theory ultimately works because it is embedded into physics, and it is so successful because it abstracts from many physical details, so that when we apply probability theory we, in most cases, are not aware of the relations to fundamental physics. 

In the above example, the sample space $\Omega$, which distinguishes the different balls, is a coarse-grained space of phase space $\mathcal{P}_B$, which describes \emph{the balls'} actual positions and velocities. These two spaces are also connected by a random variable $X_{B}: \mathcal{P}_B \rightarrow \Omega$. We can even go one floor deeper to the fundamental phase space. Every ball is a macroscopic object consisting of zillions of tiny particles. The positions and momenta of these particles are summarized in the fundamental phase space $\mathcal{P}_{f}$ . Again a random variable $X_{f}$ connects this fundamental space to $\mathcal{P}_B$, that is, $X_{f}:\mathcal{P}_{f}\rightarrow \mathcal{P}_B$. 

Of course, this interpretation of probability theory will not be shared by subjective Bayesians and other schools. My goal is not to provide a framework that suits all interpretations of probability but rather to interpret probability theory in such a way that is best suited for a modern version of frequentism. 

\subsubsection{Probability from Typicality}
\label{subsec:prob-typ}

Let us now apply all this to demonstrate how probability arises from typicality. Recall Boltzmann's explanation that we observe certain physical processes only in one direction: it is typical that ice cubes melt and not unmelt because the universe started in a low entropy macrostate, where most of the initial microstates yield a universe in which ice cubes melt. A very similar kind of explanation can be given for the emergence of probabilities from a deterministic dynamics. For simplicity's sake, I'll restrict myself to coin tosses, which is a \emph{deterministic} physical process following the laws of Newtonian mechanics, but it is,  in principle, straightforward to generalize the main idea to other physical processes and to other deterministic physical theories.    

First, there is an observational fact about coin tosses, as there is an observational fact about the thermodynamic behavior of ice cubes:  When we toss a coin thousands of times, we see that heads and tails appear approximately half the time, and the more we toss the closer the fraction of heads and tails approaches $\sfrac{1}{2}$. For instance, \citet{Kerrich:1946aa} noted to have tossed a coin 10,000 times of which heads appeared 5,067 times, and it's also said that Karl Pearson tossed a coin 24,000 times of which heads appeared 12,012 times \citep[see][although no source for Pearson's experiment is given]{Kuchenhoff:2008aa}.

Second, recall from the thermodynamic arrow of time that almost all points in phase space are in thermal equilibrium, where every point represents the physical state of a gas in a box or the entire universe. When the system starts from a low-entropy macrostate, statistical mechanics says that almost all phase space points within this macrostate will follow a trajectory according to the Newtonian laws of motion that leads to thermal equilibrium (for not too long time scales). If, say, $\Gamma$ is this low-entropy macrostate and $P$ is the property ``following a trajectory that leads to thermal equilibrium,'' then the property $P$ is said to be typical in $\Gamma$ \citep[see][p.\ 4, for this general framework]{Wilhelm:2019aa}. (We can imagine the property $P$ to give a certain color to phase-space points.) Next, let's say we are interested in the behavior of a subsystem with respect to the initial conditions of a larger system, for example, gases in a box with respect to the initial conditions of the entire universe. Then given the special low-entropy initial macrostate of the universe, it is typical (within this macrostate) that subsystems in this universe will reflect thermodynamically time-oriented behavior. In other words, $\Gamma$ would be the low-entropy initial macrostate of the entire universe, and the property $P$ would be ``subsystems reflect thermodynamically time-oriented behavior''. (Then again, almost all points in this macrostate would have the same color). 

This relation between the behavior of subsystems and the initial conditions of the universe is central to typicality frequentism.
When we apply this picture to the coin toss, we need to start with the phase space regions of the entire universe in which coins exist. The relevant property $P$ is ``long-term frequencies of fair coin-tosses are approximately \sfrac{1}{2}''. It turns out that almost all universes share this property. 

All this can be mathematically captured by the (weak) law of large numbers:\footnote{Whenever I refer to the law of large numbers, I always mean the weak law of large numbers.}
$$
\lambda \left(\left\lvert \frac{1}{N}\sum_{k=1}^{N}X_k(x)-\frac{1}{2}\right\rvert<\epsilon\right)\approx 1,
$$
where $\epsilon$ is an arbitrary small real number, $N$ is taken to be very large, the random variables $X_k$ represents the $k$th toss, $X_k(x)$ is the result of the $k$th toss determined \emph{by the initial condition of the universe $x$}, and $\lambda$ is the measure of typicality. For typical coin tosses, that is, for most universes in which coins are tossed, which translates mathematically into $\lambda(\cdot) \approx 1$, the arithmetical mean of an actual run of tosses does not deviate from $\sfrac{1}{2}$ more than $\epsilon$.  So in any sufficiently long finite series of flips in these generic universes the frequency of heads and tails will be in the range of $50\% \pm \epsilon$ for some specific $\epsilon$. 

There is something different and something similar between finite and infinite sequences. In both cases the fraction of heads and tails lies within $\pm \epsilon$ from 50\%, but $\epsilon$ in the finite case cannot be arbitrarily small and the actual frequency to be (typically) within the error bounds, whereas that is the case for infinite (or sufficiently long) sequences according to the law of large numbers. However small $\epsilon$ is chosen, it is typical that an infinite series of coin flips will have a limiting frequency within $50\% \pm \epsilon$. Note also that the finite case has to be sufficiently large in order to show some robust behavior.\footnote{\label{fn:lln-finite}
More precisely, there are three parameters in the law of large numbers that are fixed successively. First, one chooses an $\epsilon$, then a $\delta$, and then sufficiently large $N>N_0$ such that:
$$
\lambda \left(\left\lvert \frac{1}{N}\sum_{k=1}^{N}X_k(x)-\frac{1}{2}\right\rvert<\epsilon\right)>1-\delta.
$$
Such an $N_0$ exists, because according to the Chebychev inequality
$$
\lambda \left(\left\lvert \frac{1}{N}\sum_{k=1}^{N}X_k(x)-\frac{1}{2}\right\rvert<\epsilon\right)>1-\frac{1}{\epsilon^2N}.
$$

}

 Moreover, the law of large numbers mathematically says that a series of coin flips that shows 100\% heads and 0\% tails after, say, 1,000,000 flips, although physically possible, would be atypical: 
 $$
\lambda \left(\left\lvert \frac{1}{N}\sum_{k=1}^{N}X_k(x)-\frac{1}{2}\right\rvert>\epsilon\right)\approx 0.
$$
For any large $\epsilon$ you choose, there is a very small set of initial conditions that would give a long series that deviate from $50\%\pm \epsilon$.  

One may argue that the law of large numbers is a limit theorem and so doesn't say anything about finite cases, nor does it say anything about what is typical or atypical (I thank an anonymous referee for raising these concerns). One needs to distinguish between \emph{what} the limit of a sequence is and \emph{how} the sequence approaches the limit. Finding out about the right convergence behavior is, for example, a major task in functional analysis and mathematical quantum mechanics \citep{Lieb:2010aa}. Here is a simple example to illustrate this point. The three sequences $\frac{1}{n}$, $\frac{1}{\ln(n)}$, and $\frac{1}{\ln (\ln (n))}$ go to $0$ for $n\rightarrow \infty$. These sequences, however, approach the limit differently: $\frac{1}{\ln(n)}$ goes to 0 more slowly than $\frac{1}{n}$, and $\frac{1}{\ln (\ln (n))}$ even more slowly than $\frac{1}{\ln(n)}$.\footnote{E.g., $\frac{1}{1,000,000}=0.000001$, $\frac{1}{\ln(1,000,000)}\approx 0.072$, and  $\frac{1}{\ln (\ln (1,000,000))}\approx 0.38$.} If one traces a certain (standard) proof of the weak law of large numbers, one finds the formula footnote \ref{fn:lln-finite}, which can be itself proven and which tells us something about the limit behavior of \emph{finite} sequences. Then, given how typicality is defined via a measure, one can indeed rephrase the law of large numbers, as well as the limit behavior of finite sequences, in terms of typicality. I admit that this is not how the law of large numbers is standardly understood, but it is a possible, and I think consistent, way of re-interpreting what the law of large numbers says \citep[see also][]{Durr:2017aa}.

After these elaborations, we can finally define what probabilities are in typicality frequentism:
\\

\noindent\fbox{%
    \parbox{\textwidth}{%
     \emph{Definition of probability:} Some state of affairs has probability $p$, if, according to a fundamental physical theory, the physical process $X_k$ yielding this state of affairs is in principle infinitely repeatable and the instances of $X_k$ are uncorrelated such that  the frequency typically (that is, in almost all universes) approaches a unique limit $p$, that is, for all $\epsilon>0$ 
     
     $\lim\limits_{N\rightarrow \infty}\lambda \left(\left\lvert \frac{1}{N}\sum_{k=1}^{N}X_k(x)-p\right\rvert<\epsilon\right)=1.$
    }%
}
\\

This definition has several important parts that I want to comment on:\footnote{I thank an anonymous referee for raising many of the following points.}
\begin{enumerate}
\item
\emph{Definition of probability}: One may argue that what follows is not a definition but rather a sufficient condition for probabilities, because what the definition says requires certain strong idealizations that may not be met. My reply is twofold. For one, this is a definition of what probabilities are in typicality frequentism. For another, if one takes a broader view of what probabilities are in general, then this ``definition'' is indeed a sufficient (but not necessary) condition of probabilities, since I am aware that other ways of talking about probabilities differs from a frequentist account. I, therefore, advocate a pluralist theory of probabilities that complements typicality frequentism in areas where typicality frequentism does not give an account of probabilities.
\item
\emph{State of affairs:} I use ``state of affairs'' instead of ``events'' that get assigned probabilities, in order not to confuse events with the standard technical term in probability theory as a subset of the sample space or, more precisely, a member of the $\sigma$-algebra. The tossing of a coin or  a ball in roulette would be examples of  ``states of affairs''. 

\item
\emph{The fundamental physical theory:} In the ideal case, the fundamental physical theory I refer to is the Theory of Everything, the unique physical theory that correctly represents the world. Since we haven't found this theory yet, other approximately true deterministic physical theories can do the job, like Newtonian physics or the de Broglie--Bohm quantum theory. The theory needs to be approximately true in order to give rise to (at least) the right statistical pattern. Newtonian physics has been proven successful in the domain of statistical mechanics, and it is good enough for most macroscopic applications. It is also very unlikely that Newtonian physics and statistical mechanics will be completely overthrown by future physical theories. It is plausible to assume that both theories will be recovered in a kind of classical limit. A candidate for a deterministic theory on the quantum level is the de Broglie--Bohm pilot-wave theory, which also allows for extension to quantum field theory. Another deterministic quantum theory is the many-worlds theory according to Everett. My introduction of probabilities is closer to the de Broglie--Bohm theory, but also Hugh Everett III wanted to base probabilities on typicality  \citep{Barrett:2017aa}. I also think that one can generalize typicality frequentism to indeterministic theories, which would be a future project and would also require to distinguish this idea from propensities. 

\item
\emph{Uncorrelated events:} The events $X_k$ (or rather the random variables) that build up the physical process need to be uncorrelated in order to converge. Standardly, the law of large numbers requires the events of the stochastic process to be stochastically independent, which is a stronger condition than being uncorrelated. If the $X_k$'s were correlated (for example, the individual tosses of a coin), then one would be able to undermine the law of large numbers, and a unique limit may not exist. Or a limit may exist but it would not be $p$, where $p$ is technically the expectation value of $X_k$.

\item
\emph{The frequency:} The frequency that is supposed to approach the limit $p$ is the relative (finite) frequency of the the physical process $X_k$: $\frac{1}{N}\sum_{k=1}^{N}X_k(x)$. For a coin toss, for example, $X_k\in\{0,1\}$ representing when a coin lands heads $X_k=0$ or tails $X_k=1$. So, $\frac{1}{N}\sum_{k=1}^{N}X_k(x)$ counts the number of tails and divides it by how often one tossed the coin. $1-\left(\frac{1}{N}\sum_{k=1}^{N}X_k(x)\right)$ would then be the relative frequency for heads.

\item
\emph{Almost all universes:} One may think that one needs to quantify over all universes in order to determine the probabilities in our universe, and, therefore, the probabilities in our universe are also determined by what happens in other universes. The first part is correct---that one needs to quantify over all possible universes---but this doesn't mean that the probabilities here are determined by the goings-on in the other universes. Rather, one needs to compare what happens here to what happens there, and the appropriate tool for this comparison is the measure of typicality. If we are in such a world in which the assumptions of the law of large numbers hold, then the probabilities are particularly robust and regular, because most of the other universes show the same statistical pattern.\footnote{There is one caveat: even if all the assumptions of the law of large numbers were fulfilled it is still possible for a sequence to have a different limit or no limit at all; the initial conditions leading to these sequences have measure zero though.} The ``atypical'' worlds widely diverge from the typical ones and also widely diverge among themselves. There is no unifying or regular behavior to be expected in these ``atypical'' universes.

\end{enumerate}

We need to distinguish between the definition of probability and the empirical significance of this number. While the number $p$, is defined in terms of infinite sequences, which cannot be instantiated in the real world, the empirical content of this number arises from its relation to finite sequences:
\\

\noindent\fbox{%
    \parbox{\textwidth}{%
     \emph{Empirical significance of $p$:} The relative frequencies of finite sequences obey the restrictions given by the law of large numbers; that is, the observed frequencies of sufficiently long finite series typically lie in an interval $p\pm \epsilon$, where $\epsilon$ is a positive number approximately equal to $0$. 
    }%
}
\\

Let me add the following comments:\footnote{I also thank here an anonymous referee for raising these issues.}
\begin{enumerate}
\item
\emph{Status of the empirical significance of $p$:}
I take it to be a true statement about the observed relative frequencies, and that this statement follows from the definition of probabilities. It is, therefore, rather a corollary than a criterion, since a criterion would be something closer to an axiom.

\item
\emph{Sufficiently long series:}
The above definition of probabilities presupposes that the physical process is ``in principle infinitely repeatable'', but, of course, it doesn't and cannot say how often the real process is actually repeated. The probability $p$ is empirically significant because it gives bounds for the real observed (and expected) relative frequencies. It may be unsatisfactory that the real process needs to be ``sufficiently long'' without a precise numerical length. The appropriate length of the series depends on many factors of the real physical set-up and the overall physical process.     
\item
\emph{Interval $p\pm \epsilon$:}
For real world cases, one has tolerances for the relative frequencies. The question is now how robust these tolerances are. A small ``uncertainty'' of $p$ would also be consistent with the observed frequency being within the interval. First, I assume the real $p$, the one given by the true Theory of Everything, to be unique. Second,  I assume that the approximately true candidate theories that are not the Theory of Everything, like Newtonian physics or the de Broglie--Bohm theory, etc., would give $p$'s that are very similar. So in this case, there may be a tiny interval or at least a point-like spread of $p$'s. And we would have to say that there are several ''candidate probabilities''. Perhaps one of them hits the true probability; I assume, however, that these ``candidate probabilities'' are very close to the true one and for all practical purposes indistinguishable. 
\end{enumerate}

In typicality frequentism, there are actually three ideas mingled together from other interpretations of probability. The first ingredient is similar to the classical interpretation of probability, which adds up different equally probable  events according to the principle of indifference. Everything in typicality frequentism hinges on a proper way of counting that leads us to distinguish typical from atypical behavior based on big and small sets, whose elements are intrinsically ``equally likely'' to occur.  Second, the definition of probabilities in terms of a limit that cannot be carried out in the real world is reminiscent of hypothetical frequentism. Third, in order to make these typical frequencies empirically meaningful one needs to introduce tolerances for finite sequences in order to have realistic frequency bands for actual processes, but in contrast to finite frequentism probabilities in typicality frequentism are \emph{not defined} by finite sequences. 

There are two ways to undermine the long-term frequency of \sfrac{1}{2}. Either one is in one of those special universes that yield a different statistical pattern for fair coin tosses, or one were able to replicate, say, with a sophisticated tossing machine, the exact conditions in every toss. The special universes that yield atypical coin behavior may reflect all kinds of long-term coin pattern: there are initial conditions of the universe that lead to 95\% heads and also to no regular behavior at all. Because of these diverse behaviors, there is no way to put these special universes under one umbrella. It is, however, appropriate to talk of a probability of 100\% showing heads in the tossing machine example. In order to get probabilities diverging from 100\% or 0\%, physics requires significant variations in the  ways a coin is tossed  (as is realistic), and these variations in fact yield robust statistical patterns.

As \citet[][Ch.\ 5]{Gillies:2000aa} describes, the problem of connecting limiting frequencies with actual finite frequencies had been raised by de Finetti against \citet{Mises:1957aa}: 

\begin{quote}
It is often thought that these objections may be escaped by observing that
the impossibility of making the relations between probabilities and 
frequencies precise is analogous to the practical impossibility that is encountered in
all the experimental sciences of relating exactly the abstract notions of the
theory and the empirical realities. The analogy is, in my view, illusory: in
the other sciences one has a theory which asserts and predicts with certainty
and exactitude what would happen if the theory were completely exact; in
the calculus of probability it is the theory itself which obliges us to admit the
possibility of all frequencies. In the other sciences the uncertainty flows
indeed from the imperfect connection between the theory and the facts; in
our case, on the contrary, it does not have its origin in this link, but in the
body of the theory itself [\dots]. \citep[][p.\ 77]{Finetti:1937aa}
\end{quote}
The criticism against frequentism is (i) that limiting frequencies as predicted in infinite series are not observed, (ii) that there is no precise way to give an interval for the empirical frequencies, and (iii) that if an interval is given it is still possible that the actual observed frequency may lie outside this interval. Von Mises argued \citep[see][p.\ 103]{Gillies:2000aa}, as described in the first sentence of de Finetti's quote, that the problem of connecting limiting frequencies with actual frequencies is no different from connecting the idealized predictions of a scientific theory with actual observations, something ubiquitous and practically unproblematic in all of the natural sciences. To which, de Finetti replied that this analogy is invalid because a scientific theory would in principle be able to make exact predictions if it were to capture sufficiently all the relevant details of the world, whereas probability theory, even in the best case, would allow significant deviations from its predictions, both from the limiting frequencies, as well as from finite frequencies. 

I think, de Finetti's criticism of von Mises is correct, and von Mises indeed overlooked the disanalogy between probability theory and the standard application of scientific theories. The imprecision of probability theory has a different origin than the imprecision of applying scientific theories or scientific models to real world cases. The main problem for von Mises was to justify were the imprecision of his frequentism comes from. Since his theory was solely based on empirical facts, the truthmakers for the predictions of probability theory need to be empirical facts too. But how can these exceptions be empirically made true if they are rarely or never observed in the first place? 

\citet[][pp.\ 217--8]{Hajek:2009aa} makes the same argument as de Finetti when he says, ``There is no Fact  of what the Hypothetical Sequences Look Like''. He imagines a coin that is just tossed once and happens to have landed Heads. Hájek then asks about the coin:

\begin{quote}
How would it have landed if tossed infinitely many times? Never mind that---let's answer a seemingly easier question: how would it have landed on the second toss? Suppose you say "Heads". Why Heads! The coin equally could have landed Tails, so I say that it would have. We can each pound the table if we like, but we can't both be right. More than that: neither of us can be right. For to give a definite answer as to how a chancy device would behave is to misunderstand chance. \citep[][p.\ 217]{Hajek:2009aa}
\end{quote}

Again, this argument is valid for the traditional version of frequentism, but in typicality frequentism a physical theory tells us ``how the coin would have landed on the second toss''. The truthmakers for the predicted frequencies come from a physical theory, in particular, from the distributions of initial conditions of the micro-constituents of the involved physical bodies and ultimately of the entire universe itself---of course, this move would be contested by an empiricist like von Mises. Typicality frequentism, thus, explains why probability theory is intrinsically imprecise and that this imprecision cannot be improved, but at least to certain degree quantified and grounded. 

\section{Defending Typicality Frequentism}
\label{sec:defense}
Typicality Frequentism combines ideas from finite frequentism, hypothetical frequentism, and the classical interpretation of probabilities. Finite frequencies (with error bounds) describe actual outcomes of a series of a chancy process; hypothetical frequencies in terms of infinite series are used to define what probabilities are; and the principle of indifference, which is the central piece of the classical interpretation, is replaced by a measure of typicality to count events on the sample space. It seems, therefore, that the critique raised against either of these interpretations of probability is again effective to undermine typicality frequentism. The principle of indifference has been rightly dismissed when an agent is truly ignorant---although it may be successfully used for symmetry arguments \citep{Zabell:2016aa}. \citet{Hajek:1996aa,Hajek:2009aa} presents a total of 30 arguments against different versions of frequentism, 15 against finite and 15 against hypothetical frequentism, demanding that in order to rescue any kind of frequentist account  all these arguments need to be countered, where one counterargument would still leave the other 29 unanswered. I won't endeavor to reply to every single argument, because not all counterarguments are in fact counterarguments but rather characterize a frequentist's account. Instead, I will first contrast typicality frequentism with two most recent competitors, the range account and the Humean Mentaculus. Then I will counter some recent arguments raised against frequentism by \citet{Caze:2016aa}, who builds on Hájek's papers.

\subsection{The Range-Account of Probabilities}

\begin{figure}[ht]
\centering
\includegraphics[width=8cm]{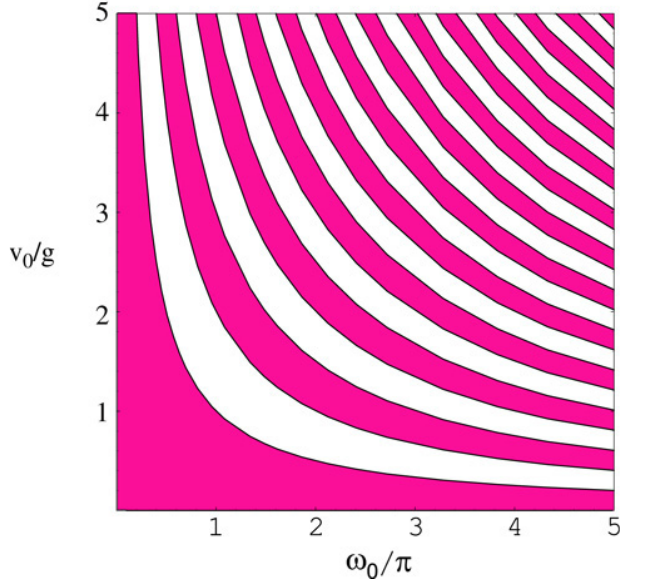}
\caption{This shows the partition of phase space for the initial conditions of a single coin, which determine how the coin will land after it is tossed. The x-axis represents the initial conditions for the angular momentum around a certain axis; the y-axis represents the vertical velocity of the entire coin. Pink areas depict the initial conditions for which the coin lands on the same face as it started, while the white areas stand for the initial conditions for which the coin changes faces. In the method of arbitrary functions, one puts a probability density on this phase space, which gives $\sfrac{1}{2}$ once integrated over all the pink or all the white areas. (Picture from \citet[][p.\ 62]{Strzako:2008aa} as an elaboration of \citet[][p.\ 193]{Keller:1986aa}.)}
\label{fig:coin-phase}
\end{figure}

The work by \citet{Kries:1886aa} was a rich source for further research. Henri Poincaré and Eduard Hopf filled in a major gap by developing \emph{the method of arbitrary functions}, which is also known as \emph{the range-account of probabilities}, advocated and further refined in different versions by \citet{Abrams:2012aa}, \citet{Rosenthal:2010aa,Rosenthal:2016aa}, and \citet{Strevens:2003aa,Strevens:2008aa,Strevens:2013ab}. Here, the probabilities, like in typicality frequentism, are related to some sort of initial conditions, but, unlike typicality frequentism, regions of phase space together with a probability density or a volume measure directly determine probabilities. For example, for the coin toss a probability density over the initial conditions for \emph{every single toss} is used (see Figure \ref{fig:coin-phase}). The physical state of a coin is completely described by its vertical velocity $v$ for the trajectory of the coin and the angular momentum $\omega$ for its rotation, given a fixed height and further simplifying restrictions, like the exclusion of bouncing \citep[see][for detailed physical models]{Keller:1986aa,Strzako:2008aa,Stefan:2017aa}. The phase space structure for the coin toss has a regular structure, in which the size of the areas leading the coin to land on the same side as it has started are approximately equal to the size of the areas for which the coin changes faces (if $v$ or $\omega$ are not too small). In the method of arbitrary functions, probabilities result when a probability density is integrated over specific regions in this phase space. The main two problems for the method of arbitrary functions is, first, to justify the particular shape of the probability density and, second, to base this justification on non-probabilistic facts in order not to explain probabilities by probabilities. This is the main point in which Abrams, Rosenthal, and Strevens disagree. They agree, however, that some measure must be used to determine the sizes of phase space regions in terms of which probabilities are defined.

The range account of probability is easily confused with typicality frequentism. First, the range account does not define probabilities in terms of frequencies. Nonetheless, Strevens's account, for example, relies on a close link to frequencies; he aims at explaining probabilities in long series of trials and facts about frequencies determine facts about the (initial) probability density \citep[see also][section 4.2 and 4.3]{Strevens:2011aa}. Second, typicality frequentism considers, like statistical mechanics, the initial conditions \emph{for the entire universe}, where a measure of typicality is imposed on. All these initial conditions are grouped into two main groups: almost all initial conditions lead to typical behavior, whereas almost no initial conditions lead to atypical behavior (there may be remaining sets that do not fit in either category, but they are not important for our current purposes). The typicality measure is only used to group the initial conditions of the universe, from which probabilities are defined in terms of frequencies.

There are several problems a range account faces, which Abrams, Rosenthal, and Strevens are aware of and have reacted to. First, one needs to justify the initial probability distribution. Where did it come from? Second, by explaining the probabilities for a coin toss by a probability distribution over the initial conditions, one would explain probabilities with probabilities. It is a challenge to explain the properties of the initial probability density from non-probabilistic facts in order not to make the theory circular. Third, a probability distribution actually contains much more information than is needed to get probabilities for frequencies. Typicality frequentism, on the other hand, introduces something weaker than an initial probability distribution that is more tailored to define probabilities as special kinds of frequencies, and it does not suffer from a circular argument (see the next section for a more detailed discussion of this point). 

Fourth, typicality frequentism can explain the initial probability distribution of the range-account (if it is the one used for frequencies). It is known that the probabilities in repeatable processes are robust under many changes of the initial probability distribution. Only very special distributions \citep[][calls them `eccentric']{Rosenthal:2016aa} would lead to different probability assignments. In typicality frequentism these  distributions are, in fact, explained to arise from special initial conditions of the universe yielding atypical behavior. \citet[][p.\ 355--6]{Strevens:2011aa} seems to be aware of this when he says, ``the typical actual pre-toss state, together with the typical actual set of timings and the typical actual coin, usually produce---because of such and such properties of the physiology of tossing---a macroperiodic set of spin speeds.'' But instead of embedding his theory into a theory of typicality, Strevens borrows from Lewis's possible-worlds semantics to explain why we observe typical frequencies in our world.


\subsection{The Humean Mentaculus}
\label{subsec:humean}

\citet[][]{Albert:2000aa,Albert:2015aa} and \citet[][]{Loewer:2001aa,Loewer:2004aa,Loewer:2012ab} have been working on a Humean account of probabilities.\footnote{\citet{Hoefer:2007aa,Hoefer:2011aa,Hoefer:2019aa} developed a more pragmatic account of Humean probabilities, which is closely linked to the Albert--Loewer account.} Similar to the range account, they postulate an initial probability distribution, but this initial probability distribution is defined on the phase space for the initial conditions of the \emph{entire} universe. More precisely, the Albert--Loewer account of probabilities consists of three postulates: 
\begin{enumerate}
\item
The fundamental deterministic laws of physics.
\item
The existence of a special (low-entropy) macrostate (called the \emph{past hypothesis}).
\item
A probability distribution over the initial conditions of the universe (within this macrostate).
\end{enumerate}
These three postulates are embedded in a Humean interpretation of laws of nature, so they are axioms in the best systematization of the Humean mosaic, balancing simplicity, strength, and fit. The initial probability distribution assigns a probability to all kinds of factual and counterfactual events. These three postulates, the \emph{Mentaculus}, are said to form a probability map of the history of the universe. Probabilities in this theory are defined, similarly to the range account, as weighed regions of initial conditions \emph{of the universe} (in phase space); in other words, one counts and weighs, according to the initial probability distribution, all possible initial conditions of the universe that would give rise to the relevant phenomenon. And again, as the range-account, the Mentaculus needs to explain what it means for the initial probability distribution to be a \emph{probability} distribution. So far, the probability distribution axiomatically introduced by the Mentaculus is merely a mathematical object that assigns numbers to certain sets.  

The central feature and goal of the Albert--Loewer account is ``to obtain a definite numerical assignment of probability to every formulable proposition about the physical history of the world'' \citep[][p.\ 7-8]{Albert:2015aa}. This probability map assigns a probability not only to coin tosses but also to events that may happen (or not) just once, like   France defending the Soccer World Cup title in 2022. There seems to be a shared intuition that these single-case probabilities are meaningful and crucial to the notion of probability---a point that has been raised against frequentism: 

\begin{quotation}
The most famous problem for finite frequentism is \emph{the problem of single case}. According to finite frequentism all single-case events automatically have the probability 0 or 1. Consider a coin that is only tossed once and comes up Heads. It appears that the probability of heads may be intermediate, but the finite frequentist is unable to say this. This goes against some strong intuitions about probability. A form of this problem remains in larger finite sequences. \citep[][p.\ 343]{Caze:2016aa}
\end{quotation}

This criticism was raised early on against frequentism, to which von Mises answered:
\begin{quote}
`The probability of winning a battle', for instance, has no place in our theory of probability, because we cannot think of a collective to which it belongs. The theory of probability cannot be applied to this problem any more than the physical concept of work can be applied to the calculation of the `work' done by an actor in reciting his part in a play. \citep[von Mises, quoted in][p.\ 98]{Gillies:2000aa}
\end{quote}
For many it was a shortcoming of frequentism that it does not assign probabilities to single events, although it ought to do so \citep[][p.\ 227--8]{Hajek:2009aa}. Von Mises argues, and I agree with him here, that scientific concepts may not capture the full range of intuitive notions and it may not even be the goal of science to form concepts that capture all the different meanings of an intuitive notion. Scientific concepts are defined in a precise way for the price of being less general. Probability, according to von Mises, is like the word ``work'' in physics, which has a precise meaning in terms of an integral of the forces along a certain path and which, thus, differs from the everyday meaning of ``work''. Von Mises was, therefore, open to a pluralistic account of probability dependent on the field of application. 

In contrast to von Mises, other frequentists tried to generalize probabilities to single cases as a kind of fictitious value:
\begin{quotation}
Frequentists from Venn to Reichenbach have attempted to show how the frequency concept can be made to apply to the single case. According to Reichenbach, the probability concept is extended by giving probability a ``fictitious'' meaning in reference to single events. We find the probability associated with an infinite sequence and transfer that value to a given single member of it. \citep[][p.\ 90]{Salmon:1966aa}
\end{quotation}
Although one can formally or ``fictitiously'' assign these numbers from frequencies to single events, their meaning is unclear, especially their meaning as something objective or physical.  This is not only a problem for frequentism, but also for the Humean Mentaculus because it is unclear what a probability in an objective or physical sense for a single event is in the first place. A purely subjective account, on the other hand, would not have this problem, as probabilities are an agent's degree of belief, which are meaningful for single events, because they capture how confident an agent is to believe a proposition. 

In the Mentaculus, probabilities are introduced by a probability density over the initial conditions of the universe, but this probability density, it shall be noted, merely axiomatically introduces numbers on the Humean mosaic. To make this distribution of numbers a \emph{probability} distribution requires further elaboration and an interpretation that turns these numbers into probabilities. This is accomplished in two steps \citep[][]{Loewer:2004aa}. First, the concept of ``fit'' is introduced. Every (probabilistic) proposition is said to have a certain degree of fit, that is, how likely it is to be true, and this is quantified by a probability. If a proposition with high probability matches the actual facts, it has a better fit than a proposition with low probability.\footnote{The concept of fit leads to the \emph{zero-fit} problem; \citet{Elga:2004aa} proposes a solution by invoking a certain notion of typicality.} Second, in order that fit in terms of probabilities is informative, an agent needs to constrain her belief according to these probabilities \citep[][p.\ 1122]{Loewer:2004aa}, and this is done according to another axiom, the \emph{Principal Principle}. It roughly says that an agent ought to adjust her degree of belief or her credence according to the probability of the proposition given by the Humean best system. 

It is not immediately clear what the physical meaning of single-case probabilities is in this Humean theory. Let us say  that there are two coins, and the Mentaculus assigns a probability of landing heads of 0.4 to one coin and 0.6 to the other. Each coin is just once tossed and then destroyed. What can these numbers 0.4 and 0.6 mean? These probabilities indeed influence, by the Principal Principle, an agent's attitude and behavior toward the outcome of the coin tosses. For example, an agent will bet differently on an event with probability of 0.4 than on an event with a probability of 0.6. It seems, however, that these single-case probabilities need also to say something about the physical events themselves, whether their occurrence is in some way constrained or not, which is then the basis for an agent to adjust her degree of belief. Moreover, this example of two coins being tossed just once is in principle repeatable, and so Humeans need to clarify the relationship between these single-case probabilities and the frequencies of repeatable coin tosses. Although the Albert--Loewer account of Humean probabilities explicitly introduces and endorses single-case probabilities, it is, as of now, unclear what their objective physical meaning is supposed to be.

\subsection{Countering Standard Critique of Frequentism}

Building on Hájek's critique of frequentism, \cite{Caze:2016aa} launched another comprehensive attack. Here, I reply to four of La Caze's arguments: (i) that frequentism is a poor \emph{analysis} of probabilities, (ii) the problem of ascertainability, (iii) the reference class problem, and (iv) that frequentism is not completely objective.
 
\subsubsection{It's a Poor \emph{Analysis} of Probabilities}
\citet{Caze:2016aa} claims that hypothetical frequencies are not the right description of probabilities because they provide a poor analysis of what probabilities are:

\begin{quotation}
The hypothetical frequentist provides an answer to the question ``What is probability?'' with an analysis that has little relationship with what most people \emph{mean} by the probability statements they make. [\dots] When stating that a specific coin has the probability of Heads of half, people are typically referring to their beliefs about the coin, their experience with this coin in a finite series of tosses, or their experience with similar-seeming coins. \citep[][p.\ 350]{Caze:2016aa}
\end{quotation}
The aim of typicality frequentism is not to reduce all ways in which probabilities are invoked to typical long-term frequencies. It, rather, aims at showing how one can derive from fundamental physics physically meaningful probabilities, and it is open to be complemented by other accounts of probability outside its scope. Given the myriads of different cases in which probabilities are used, it is plausible that all these cases are not unified by one account. Typicality frequentism would be, in my view, one piece in a pluralistic landscape of probabilities. Moreover, if typicality frequentism is true, then people may need to re-think their intuitions they have about probabilities of coins and other physical processes. I aim at giving an account of objective probability, but I agree that we also need an account of subjective probabilities, and I can envision that it may be possible, in certain circumstances, to connect a particular interpretation of subjective probabilities with typicality frequentism. 


\subsubsection{The Problem of Ascertainability}

``[T]he problem of ascertainability is the most fundamental difficulty the frequency interpretation faces,'' says \citep[][p.\ 89--90]{Salmon:1966aa}, and he defines this problem in the following way:

\begin{quotation}
\emph{Ascertainability.} This criterion requires that there be some method by which, \emph{in principle at least}, we can ascertain values of probabilities. It merely expresses the fact that a concept of probability will be useless if it is impossible \emph{in principle} to find out what the probabilities are. \citep[][p.\ 64, my emphasis]{Salmon:1966aa}
\end{quotation}
Actually, all interpretations of probability face in one form or other the problem of ascertainability, that is, how to assign probabilities in practice. A meaningful definition is not enough, because it may lack the instructions for how to pick the right probabilities. Salmon stresses that these instructions, however, are supposed to be applicable only \emph{in principle}, and not necessarily in actual practice. Applied to hypothetical frequencies, they are said to be unascertainable for the following reasons \citep[see also][pp.\ 214--5]{Hajek:2009aa}:

\begin{quotation}
To ascertain a hypothetical frequency with certainty we would need to observe an infinite number of trials. Assuming that a specific sequence of observations will converge to a limiting relative frequency, there is no guarantee that it will do so within the number of trials that will be observed. And if a relative frequency appears to have converged in a finite number of trials, it is always possible that the relative frequency diverges from this value in subsequent trials. These points are direct consequences of the mathematics of infinite sequences. The task for the frequentist is to justify inferring a (frequentist) probability from a relative frequency observed in a finite number of trials, and there is no deductively valid way to do this. \citep[][p.\ 353]{Caze:2016aa}
\end{quotation}
The argument amounts to the correct observation that we cannot figure out the true probability (as a hypothetical frequency) by observing finite frequencies. This objection is particularly damaging to von Mises and Reichenbach because they defined the probabilities in the spirit of logical empiricism based on obersvation. The only means that they had to reach the hypothetical frequencies is by means of observable finite frequencies. In order to mitigate this problem, von Mises introduced two principles \citep[][p.\ 100--1]{Rowbottom:2015aa}:
\begin{enumerate}
\item
\emph{Law of Stability}: The relative frequencies of attributes in collectives become increasingly stable as observations increase. 
\item
\emph{Law of Randomness}: Collectives involve random sequences, in the sense that they contain no predictable patterns of attributes. 
\end{enumerate}
These laws are arguably ad hoc in von Mises theory, but at least they may be justified by induction.

Similarly to von Mises, \citet{Reichenbach:1949aa} bridged the gap between finite and infinite sequences by induction; Reichenbach called his law the \emph{Rule of Induction by Enumeration}. Starting with an infinite sequence of events $A$, we are interested in the relative frequency that some feature $B$ occurs in this sequence. We can only observe a finite sequence of events of length $n$, for example. The frequency of feature $B$ among the first $n$ members of $A$ is, say, $F^n(A,B)=\sfrac{m}{n}$. In order to infer the limiting frequency, the \emph{Rule of Induction by Enumeration} needs to be applied: Given $F^n(A,B)=\sfrac{m}{n}$, to infer that $\lim\limits_{n\rightarrow \infty} F^n(A,B)=\sfrac{m}{n}$ \citep[][pp.\ 85--6]{Salmon:1966aa}. 


La Caze, on the other hand, demands a deductive way to get to the hypothetical frequencies, and this can be, in principle, accomplished by typicality frequentism, as the hypothetical frequencies are predictions of the laws of physics about typical behavior. By applying a physical theory, probably by building a model as is standard in many cases \citep{Cartwright:1983aa,Morgan:1999aa,Giere:2004aa}, the probabilities fall out of the theory as any other empirical prediction. This move was not possible for von Mises and Reichenbach, as they based their probabilities on observable behavior of the physical processes. Typicality frequentism adheres to Salmon's requirement for solving the problem of ascertainability, because we can access the information of a physical theory \emph{in principle}; in practice, there might be strong limitations on how to access all this information, but these obstacles are not of a different nature than we normally encounter in other kinds of empirical predictions. 

One may argue that the ``in principle'' in typicality frequentism does a lot of work.\footnote{Thanks to an anonymous referee for raising this point.} If we have a powerful enough physical theory that also makes it easy to extract empirical predictions, then we would be able to solve the problem of ascertainability. But what if we cannot extract this information from a physical theory (for whatever reason)? Then either we need to extract the right frequencies from observations, or we need to apply further metaphysical or physical assumptions. Both paths are problematic: the first because we would fall back to the (empirical) problem of hypothetical frequentism, the second because further theoretical assumptions need to be justified. I grant that this is argument poses a challenge to the epistemology of typicality frequentism, that is, how to ascertain the probabilities in practice. It is in general very hard, and mostly impossible, to extract precise empirical information from a physical theory for sufficiently complex systems---we cannot even analytically solve the three-body problem in classical physics. Therefore, for practical purposes we rely on other means to make empirical predictions: for example, by making certain idealizations and approximations. In the case of probabilities, we may need to rely on past incomplete empirical observations for future predictions, or we may use theoretical assumptions, like symmetry arguments \citep{Zabell:2016aa}.

\subsubsection{The Reference Class Problem}

The reference class problem is generally regarded to sound another death knell to frequentism \citep[][pp.\ 219]{Hajek:2009aa}, although it was originally raised \emph{by frequentists}, like \citet{Venn:1888aa} and \citet{Reichenbach:1949aa}, \emph{against single-case probabilities}:
\begin{quote}
If we are asked to find the probability holding for an individual future event,
we must first incorporate the case in a suitable reference class. An individual
thing or event may be incorporated in many reference classes, from which different probabilities will result. This ambiguity has been called the \emph{problem of the
reference class}. \citep[][p.\ 374]{Reichenbach:1949aa}
\end{quote}
The reference class problem for single-case probabilities is a problem of how to get the probability of one event when it can be part of many collections. Venn's example is the single-case probability of a man called John Smith, aged 50, to die at age 61. In order to make a qualified prediction of Smith's life in the future eleven years, one needs to compare Smith with other people similar to Smith. In order to extract single-case probabilities from frequentism, one would need to find a set of people similar to Smith and  who live until 61 and compare this number with all the people of this age. The problem is, however, that it is not clear which properties the reference class, that is, the people similar to Smith, need to have in order to count as ''similar to Smith,'' (also because Smith himself has so many different properties).

This example can be transferred into a reference class problem for frequentism in general \citep[][section 16.4.4]{Caze:2016aa}; we just need to add to John Smith any finite number of people of the same age and ask about the probability of their life expectancy until age 61. What is the correct infinite collection of people that give rise to the right probability? More precisely, given a finite sequence of events $(x_1,\dots,x_n)$, what is the appropriate infinite sequence $(y_1,y_2,\dots)$ that we shall associate with $(x_1,\dots,x_n)$ in order to assign the probability $p=\lim\limits_{m\rightarrow\infty}\frac{1}{m}\sum_{i=1}^{m}y_i$ for some feature of $(x_1,\dots,x_n)$?

Furthermore, having found a suitable (or even the ``correct'?) reference class, the order of the members of the reference class may change the probability, and there may be even an ordering where the sequence does not converge and no probability can be assigned in the first place. \citet[][p.\ 567]{Hajek:2007aa} calls this subcategory of the reference class problem the \emph{reference sequence problem}. Von Mises dealt with the reference sequence problem by restricting the admissible sequences to give a unique ordering; these sequences, he called \emph{collectives}, and they are defined by means of his two laws of probability, the law of stability and the law of randomness. With this move, von Mises could only solve, or propose a solution to, one aspect of the reference class problem, namely, what \citet[][p.\ 565 ]{Hajek:2007aa} calls the \emph{metaphysical} reference class problem. Given the two laws of probability, there is (hopefully) a fact what the correct reference class is and what accordingly the probability is. Still, this information may be practically inaccessible for an agent, which amounts to an \emph{epistemological} reference class problem.

Does the reference class problem only arise in frequentist interpretations of probabilities? \citet{Hajek:2007aa} argues that basically \emph{all} interpretations of probability face their version of the reference class problem, and the best we can hope for is to solve the metaphysical problem---the epistemological problem will always remain. And theories that do not face a reference class problem in the first place, like radical subjectivists à la de Finetti or certain versions of the propensity interpretation, are, according to Hájek, no-theory theories of probability, because they do not sufficiently specify what probabilities are and how they are to be used to guide agents' beliefs and actions.

Typicality frequentism indeed solves the metaphysical reference class problem by means of a physical theory, something Salmon also mentioned as a way out: 
\begin{quotation}
When a sequence is generated by a physical process that is well understood in terms of accepted physical theory, we may be able to make theoretical inferences concerning convergence properties. For instance, our present knowledge of mechanics enables us to infer the frequency behavior of many kinds of gambling mechanisms. Our theory of probability must allow room for inferences of this kind. The basic problem, however, concerns sequences of events for which we are lacking such physical knowledge.  \citep[][p.\ 84]{Salmon:1966aa}
\end{quotation}
The Theory of Everything ultimately determines the underlying physical processes of a random sequence, and thus determines the limit of a finite sequence if one were to repeat it infinitely. The reference class problem is solved in typicality frequentism, because the reference class is the finite sequence itself which gets extrapolated into an infinite sequence by means of the Theory of Everything. Since we do not yet have a Theory of Everything, any candidate for a fundamental physical theory determines the behavior of the reference class. In other words, the truthmaker for singling out the reference class and the corresponding behavior is the Theory of Everything, and for the current situation we can replace the Theory of Everything by an appropriate candidate for a fundamental physical theory or by a model of the physical theory (given certain idealizations). So the gap in Reichenbach's Rule of Induction by Enumeration is closed not by induction from the observable sequence itself, but by a physical theory describing the physical processes underlying the sequence. 

In a similar vein, the reference sequence problem is tackled. Intricate orderings that yield different limits or no limit at all are physically possible but atypical, given the initial conditions of the universe, which determine the physical conditions of the physical processes governing the sequence.\footnote{Something similar has been proposed in certain versions of the propensity interpretation. 
\citet[][p.\ 56]{Miller:1994aa} says that propensities are determined by ``the complete situation of the universe (or the light-cone) at the time, and, for \citet[][p.\ 195]{Fetzer:1982aa}, they are determined  by ``a complete set of (nomically and/or causally) relevant conditions [\dots] which happens to be instantiated in that world at that time.'' These solutions, however, are not satisfactory for \citet[][p.\ 576]{Hajek:2007aa} because the propensities, such defined, are not accessible to an agent to assign probabilities in practice. Therefore, he subsumes these proposals under no-theory theories of probabilities.} More precisely, there is a physically distinguished ``natural'' ordering of the sequence, namely, the temporal ordering as determined or predicted by physics.  
\citet[][p.\ 111]{Rowbottom:2015aa} presents an argument that physics is not able to single out a natural order for sequences , because, according to special relativity, the order of, say, coin flips depends on the state of motion of an observer. So two observers on two different trajectories may disagree on the order of the same coin flips that they observe. But this would be only correct when the observers would see \emph{two} different sequences of coin flips that are space-like separated. If Rowbottom refers to one sequence of coin flips, and I assume he does because this is the relevant case at issue, then the coin flips are time-like separated, and, according to special relativity, the temporal order of time-like separated events are objective, that is, independent of the state of motion of observers.





\subsubsection{Frequentism Is Not Completely Objective}
Von Mises \citeyearpar[][p.\ 14]{Mises:1957aa} makes a strong assertion about frequentistic probabilities when he says, ''The probability of a 6 is a physical property of a given die and is a property analogous to its mass, specific heat, or electrical resistance.'' I agree with \citet[][section 16.4.5]{Caze:2016aa} that frequentism does not oblige upon us this strong metaphysical interpretation of probabilities, but I disagree that frequentistic probabilities are not objective. For La Caze, ``[h]ypothetical frequencies are not divorced from consideration of personal factors (including beliefs).'' 

His argument goes like this. Since the main advantage proclaimed of frequentism is that it introduces objective probabilities, any subjective trace in frequentistic probabilities would undermine the entire project. The subjectivity that frequentism relies on comes from how the particular physical process that gives rise to frequencies is modeled. The probabilities for a coin toss, for example, depend on how the properties of the coin and the tossing mechanism are modeled. That some particular physical model is suitable for giving rise to the proper frequencies needs to be judged by an agent. And this judgement is unequivocally subjective, as \citet[][p.\ 358]{Caze:2016aa} says, ``Scientists employing frequentists probabilities need to make a judgement that the data-generating processes providing the measured outcomes of the study are adequately modeled by one or more of these approaches to specifying the requirements on the expected sequence of outcomes.'' The bar raised by this requirement is so high that basically all our physical predictions are deemed to be subjective, because they depend on certain idealizations to be made by an agent. The practice of physics has for practical matters this ``subjective'' ingredient but it does not make physics a subjective science. Therefore, I do not see that frequentistic probabilities are less objective than other predictions of physics. 

\citet[][pp.\ 215--7]{Hajek:2009aa} also criticizes the idealizations made in hypothetical frequentism, but he approaches this problem from a different direction:

\begin{quote}
Consider the coin toss. We are supposed to imagine infinitely many results of tossing the coin: that is, a world in which coins are `immortal', lasting forever, coin-tossers are immortal and never tire of tossing (or something similar anyway), or else in which coin tosses can be completed in ever shorter intervals of time... In short, we are supposed to imagine \emph{utterly bizarre} worlds [\dots]. \citep[][pp.\ 215--6]{Hajek:2009aa} 
\end{quote}
For Hájek, the problem of hypothetical frequentism lies in the definition of probabilities: in order to define hypothetical frequencies ``utterly bizarre'' counterfactual scenarios need to be set up that ``would have to be \emph{very} different from the actual world'' \citep[][pp.\ 215]{Hajek:2009aa}. I think this problem can be remedied by a physical theory and the laws of nature in such a theory. We know that laws of nature ground facts beyond the actual regularities \citep[e.g.,][]{Maudlin:2007ah}. The counterfactual idealizations that need to be made for hypothetical frequentism---and also for typicality frequentism---may be more radical or more detached from the actual world than in other applications, like in the normal way of model building \citep{Morgan:1999aa}, but they can be still grounded and made true by the laws in a physical theory.

\subsection{Objections and Replies}
\begin{enumerate}
\item
\emph{Typicality seems to be too vague. How can it be meaningful?}

Typicality is intentionally a vague term. Not all notions need to be precise to be meaningful. We know when someone is tall or when someone is bald. Of course, there may be borderline cases when we may debate is this person really tall or really bald, but for all practical purposes there is no ambiguity. The same we encounter in physics. The initial macrostate the universe evolved from according to statistical mechanics  is also vague, because the boundaries are fuzzy and not precisely specified. But when we reason about the evolution of the universe we talk about microstates that do not lie on the boundary, so this vagueness is harmless. It is a strength of the notion of typicality to be vague, because we don't need to cope with unnecessary details in our explanation and we can use typicality in many different areas.
\item
\emph{What is a formal definition of typicality?}

In many cases, typicality does not need a formal definition. It is basically a technical term for \emph{most} or \emph{almost all}. \citet{Maudlin:2019ab} and \citet{Wilhelm:2019aa} propose two different approaches to formalize typicality. Maudlin interprets typicality as a second-order predicate, that is, a predicate of a predicate. We formally write $F(X)$ symbolizing that $X$ has property $F$. Typicality would be a further qualification between $X$ and $F$. $T(F,X)$ would symbolize that it is typical for $X$ to have $F$. One may even consider typicality as another quantifier. Wilhelm, on the other hand, focuses on the explanatory scheme of typicality explanations and points out that it resembles Hempel's deductive–nomological model.  
\item
\emph{What is the relationship between a probability measure and a typicality measure?}

Mathematically, a typicality measure  is usually represented as a probability measure, but a probability measure contains more information than is actually needed:

\begin{quotation}
While typicality is usually defined -- as it was here -- in terms of a probability measure, the basic concept is not genuinely probabilistic, but rather a less detailed concept. A measure $\mu$ of typicality need not be countably additive, nor even finitely additive. Moreover, for any event $E$, if $\mu$ is merely a measure of typicality, there is no point worrying about, nor any sense to, the question  as to the real meaning of say `$\mu(E)=\sfrac{1}{2}$`. Distinctions such as between `$\mu(E)=\sfrac{1}{2}$` and `$\mu(E)=\sfrac{3}{4}$`
are distinctions without a difference.

The only thing that matters for a measure $\mu$ of typicality is `$\mu(E)\ll1$´: a measure of typicality plays solely the role of informing us when a set $E$ of exceptions is sufficiently small that we may in effect ignore it and regard the phenomenon in question occurring of the set $E$, as having been explained. \citep[][p.\ 15]{Goldstein:2001aa}
\end{quotation}

\item
\emph{What is the difference between typicality and probability? Is \emph{typical} just another word for \emph{highly probable} and \emph{atypical} for \emph{highly improbable}?}

Historically typicality evolved from abstracting from highly probable cases. Boltzmann, for example, said that the second law of thermodynamics makes it highly probable that a gas in a box equilibrates. But I think that typicality is a more primitive notion than and different from probability, and this paper showed how one can reduce probabilities to typicality. Typicality is a much less-fine grained and more general concept than probability. 
\item
\emph{There are cases where something typical is highly improbable. For example, a long, well-mixed sequence of heads is typical but improbable, e.g. HTHTTHHHTTTHTHHT. Or a very specific event may be typical but improbable, e.g. the probability of randomly selecting from the US population a man of height exactly 175.4 cm is very low, even though this is the average height, and in a good sense typical. How can one reconcile that?}\footnote{Thanks to an anonymous referee for raising this issue, who I quote almost verbatim.}

The difference between typicality and probability have been addressed in more detail in \citet{Wilhelm:2019aa}. Typicality is not a categorical property. So it doesn't make sense to say that something is typical by itself. There needs to be always a reference: “typical with respect to what?” It is typical for clovers to have three leaves, because in the class of all clovers most of them have three leaves. If we zoom in too much, for instance, when comparing the particular shapes of the leaves, every leaf may be unique, and we may not be able to find any “typical shape”. Applied to the coin toss, if we zoom in the particular pattern of a series of tosses, and ask, “Is HTHTTHHHTTTHTHHT typical?”. The right answer is, “Typical with respect to what?” Typical with respect to the number of heads and tails. Then yes, because approximately 50\% are heads and tails (I ignore that the series needs to be much longer to make such a statement). But what about the particular pattern HTHTTHHHTTTHTHHT? It is very unlikely to repeat this particular pattern in an actual coin toss. But this is the case for any particular pattern. The same is true in statistical mechanics: every particular trajectory in phase space has measure zero and is therefore very unlikely (or atypical, although it is meaningless to talk about atypicality per se too). This point has been raised against typicality by \citet{Uffink:2007aa}, and I think it has been rightly answered by \citet[][section 5.2]{Lazarovici:2015aa}. 
I agree on the example of the height, which is similar to the way I define probabilities. The actual height 175.4 cm is rarely found in a person but most people are close to the average.

\item
\emph{If you reduce probabilities to typical longterm frequencies, then you cannot account for all the uses of probability. Especially single-case probabilities lack an explanation.}

That is correct, but I claim that single-case probabilities are not meaningfully interpreted as some kind of frequency. They may be properly construed as purely subjective degrees of belief, as a kind of tool in Bayesian updating, but not in an ontological sense. Therefore, I endorse a pluralistic account of probabilities tailored to different applications.
\end{enumerate}

\section{Conclusion}
If our world is correctly described by a deterministic physical theory, then every event is determined by the initial conditions of the universe. Typicality frequentism builds on this insight and singles out physical processes that give rise to stable long-term frequencies. If these frequencies are typical they define probabilities. As I have shown, the essential idea behind this approach comes from how Boltzmann explained the thermodynamic arrow of time and how he reduced thermodynamics to statistical mechanics. The main advantage I see with typicality frequentism is that it carves objective probabilities at the right joint by specifying those kinds of probabilities that are meaningful within physics. In this way, typicality frequentism does not face the same problems as traditional empiricist accounts of frequentism do. Other applications of probability beyond physics may be properly described by subjective approaches that would complement to a pluralistic picture of probabilities. 


\section*{Acknowledgements}
I wish to thank Frederick Eberhardt, Christopher Hitchcock,  and Charles Sebens for their helpful and detailed comments on previous drafts of this paper. I also wish to thank David Albert, Jeffrey Barrett, Detlef Dürr,  Sheldon Goldstein,  Dustin Lazarovici, Barry Loewer, Tim Maudlin,  and Isaac Wilhelm for many invaluable hours of discussions. I also thank the members of the \emph{Caltech Philosophy of Physics Reading Group}, in particular Joshua Eisentahl and James Woodward. I want to thank two anonymous reviewers for their helpful comments, which significantly improved the paper.  Especially one of the anonymous reviewers spent considerable time and effort in the review process; I particularly thank this reviewer.

\bibliographystyle{abbrvnat}
\bibliography{references}
\end{document}